\documentclass[aps,prl,showpacs,showkeys,reprint,superscriptaddress]{revtex4-1}
\usepackage{graphicx}
\usepackage{bm}
\usepackage{braket}
\usepackage[normalem]{ulem}
\usepackage{amsmath}
\usepackage{mathrsfs}
\usepackage{multirow}
\usepackage{rotating}
\usepackage{todonotes}
\usepackage{array}
\usepackage{color}
\usepackage{feynmf}
\usepackage[colorlinks=true,linkcolor=blue, citecolor=blue, urlcolor=blue]{hyperref}
\usepackage{color,soul}
\usepackage{mathtools}

\newcolumntype{L}[1]{>{\raggedright\let\newline\\\arraybackslash\hspace{0pt}}m{#1}}
\newcolumntype{C}[1]{>{\centering\let\newline\\\arraybackslash\hspace{0pt}}m{#1}}
\newcolumntype{R}[1]{>{\raggedleft\let\newline\\\arraybackslash\hspace{0pt}}m{#1}}

\renewcommand{\thispagestyle}[1]{}

\begin{document}
\title{Mode specific electronic friction in dissociative chemisorption on metal surfaces: H$_2$ on Ag(111)}
\begin{flushright} accepted for publication in Phys. Rev. Lett. \end{flushright}
\author{Reinhard J. Maurer}
\email{reinhard.maurer@yale.edu}
\affiliation{Department of Chemistry, Yale University, New Haven, CT 06520, USA}
\author{Bin Jiang}
\affiliation{Department of Chemical Physics, University of Science \& Technology of China, Hefei, Anhui 230026, China}
\author{Hua Guo}
\affiliation{Department of Chemistry and Chemical Biology, University of New Mexico, Albuquerque, NM 87131, USA}
\author{John C. Tully}
\email{john.tully@yale.edu}
\affiliation{Department of Chemistry, Yale University, New Haven, CT 06520, USA}

\date{\today}

\begin{abstract}
Electronic friction and the ensuing nonadiabatic energy loss play an important role in chemical reaction dynamics at metal surfaces. Using molecular dynamics with electronic friction evaluated on-the-fly from Density Functional Theory, we find strong mode dependence and a dominance of nonadiabatic energy loss along the bond stretch coordinate for scattering and dissociative chemisorption of H$_2$ on the Ag(111) surface. Exemplary trajectories with varying initial conditions indicate that this mode-specificity translates into modulated energy loss during a dissociative chemisorption event. Despite minor  nonadiabatic energy loss of about 5\%, the directionality of friction forces induces dynamical steering that affects individual reaction outcomes, specifically for low-incidence energies and vibrationally excited molecules. Mode-specific friction induces enhanced loss of rovibrational rather than translational energy and will be most visible in its effect on final energy distributions in molecular scattering experiments. 
\end{abstract}


\maketitle

The reaction dynamics of elementary processes in heterogeneous catalysis depend sensitively on the energy exchange between adsorbate and substrate degrees of freedom, including substrate phonons and electrons. Specifically in the case of chemical reactions on metal surfaces, nonadiabatic adsorbate-substrate energy exchange \emph{via} electron-hole pair excitations (EHPs) has been a focus of interest in literature~\cite{Wodtke2004, Hasselbrink2009, Shenvi2009a}. Recent experiments of atomic hydrogen scattering on Au(111) gave evidence of inelastic scattering due to EHP-induced energy loss.~\cite{Bunermann2015} The ensuing nonadiabatic force acting on the adsorbate and mediating this energy loss is often referred to as electronic friction. Electric currents, induced by electronic friction have been measured for a variety of chemical reactions on surfaces~\cite{Gergen2001,Hasselbrink2009, Schindler2013} and a selective reaction control theory coined "hot-electron chemistry" has emerged from the idea of selectively funneling energy into reaction channels via EHPs~\cite{Ji2005, Mukherjee2013, Park2015, Kim2016}.

Despite the experimental evidence, purely adiabatic theories have reproduced experimental data with reasonable accuracy for molecular hydrogen scattering and dissociation on substrates such as Cu(111)~\cite{Diaz2009} and Ru(0001)~\cite{Fuchsel2013}. From this indirect evidence, it has often been argued that EHPs can be safely neglected in studying scattering probabilities and molecular motion leading up to dissociation~\cite{Kroes2016}. On the contrary, a recent simulation study has identified EHPs as the dominant energy loss channel after dissociation. Here, the energy loss through EHPs was found to exceed phonon energy loss by a factor of five~\cite{Blanco-Rey2014}.

Much less is known about the role of nonadiabatic effects near the transition state of dissociative chemisorption (DC). Several first-principles simulation studies~\cite{Juaristi2008, Fuchsel2013, Jiang2016} have argued that EHP effects are limited in DC, because the interaction time between the molecule and surface is short. On the other hand, Luntz \emph{et al.}~\cite{Luntz2009,Luntz2005} have argued that established models to account for electronic friction based on the local electron density, termed local density friction approximation (LDFA),~\cite{Li1992, Echenique1986} neglect the molecular electronic structure of the adsorbate and misrepresent friction during chemical transformations. The majority of recent studies have employed friction models based on LDFA, assuming that frictional effects on H$_2$ are isotropic.~\cite{Juaristi2008} Electronic friction models based on first-order time-dependent perturbation theory (TDPT),~\cite{Hellsing1984, Head-Gordon1992, Trail2002, Forsblom2007, Meyer2011, Maurer2016a} that account for the adsorbate molecular structure, have revealed significant levels of anisotropy and mode dependence in vibrational energy loss of diatomics on metal surfaces~\cite{Askerka2016, Maurer2016a}. This mode dependence is visible in experiment.~\cite{Inoue2016}, with recent molecular scattering experiments adding new evidence for its experimental relevance.~\cite{Shuai2017}  Unfortunately, the computational cost of TDPT models hitherto did not allow their routine  application in full-dimensional molecular dynamics simulations (MD), leaving questions relating to EHP-induced effects on dynamical steering or mode-specific energy redistribution unanswered. 

In this Letter, we combine MD with an efficient implementation of tensorial electronic friction based on TDPT to describe electronic friction effects during chemisorption of molecular hydrogen on a close-packed silver surface. This allows us, for the first time, to overcome restrictions of previous methods by accounting for adsorbate molecular structure, anisotropy, and mode coupling within full-dimensional MD. We select H$_2$ on Ag(111) as a prototypical example of DC, because EHPs are believed to play a larger role on coinage metal surfaces, where the dissociation barrier lies closer to the surface.~\cite{Kroes2016, Wijzenbroek2016} Indeed, we find electronic friction to act dominantly along the intramolecular stretch motion as the molecule dissociates. As a consequence, electronic friction induces modulated energy loss during scattering events. Whereas reaction outcomes of high-energy molecules are not sensitive to EHP effects, conditions of low incident translational energy or initial vibrational excitation are affected in their reaction outcomes and final energy distributions. The strong frictional mode specificity along the intramolecular stretch is visible in a more effective loss of rovibrational energy than translational energy, serving as a potential experimental signature of mode-specific nonadiabatic effects.~\cite{Shuai2017}

\begin{figure}
 \centering\includegraphics[width=1.0\columnwidth]{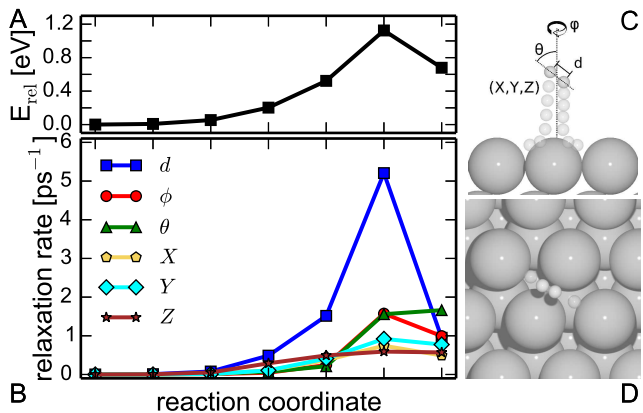}
 \caption{\label{fig-neb-friction} (A) Relative energy along a minimum energy path (MEP) of H$_2$ dissociation on Ag(111). (B) 
 Nonadiabatic relaxation rates in ps$^{-1}$ obtained from the diagonal elements of electronic friction tensor in internal coordinates: bond stretch $d$, azimuthal angle ($\phi$), polar angle ($\theta$), and the three Cartesian center of mass coordinates of the Hydrogen molecule (X,Y,Z). Off-diagonal elements (not shown) can modify relaxation rates by 30\% or more. (C) Side view of the dissociation along the minimum energy path. (D) Top view of start and end point of MEP.}
\end{figure}

Electronic friction originates from the coupling of electronic excitations with adsorbate nuclear motion. This effect can be captured within mixed quantum-classical theories.~\cite{Dou2016a,Ryabinkin2017} The simplest version uses a Langevin expression:~\cite{Head-Gordon1995} 
\begin{align}\label{eq-langevin}
M\ddot{R_i}= - \frac{\partial{V\bm{(R)}}}{\partial{R_i}} \overbrace{- \sum_{j} \Lambda_{ij}\dot{R_j}}^{F_{\mathrm{damp},i}} + \mathscr{R}_i(t).
\end{align}
Atomic positions $R_i$ evolve due to forces from an \emph{ab-initio} potential energy surface (PES), $V\mathbf{(R)}$, and frictional forces due to EHPs are captured by an electronic friction tensor $\mathbf{\Lambda}$ and thermal white noise $\mathscr{R}_i(t)$. The latter ensures detailed balance~\cite{Kubo2012}. In eq.~\ref{eq-langevin}, $\mathbf{\Lambda}$ is a (3N$\times$3N) matrix for N atoms (in our case 2 hydrogen atoms), wherein each element corresponds to the nonadiabatic relaxation rate due to adsorbate motion along that coordinate or, in the case of off-diagonal elements, the coupling between two coordinates.
We calculate nonadiabatic relaxation rates using TDPT within the constant-coupling approximation required by eq.~\ref{eq-langevin}, where we average over EHP excitations within the relevant energy regime using single particle states from Density Functional Theory (DFT) and a Gaussian envelope of width 0.6~eV~\cite{Maurer2016a,Askerka2016}. This method is implemented in the all-electron code FHI-aims, which employs numerical atomic-orbitals.~\cite{Blum2009} We use the exchange-correlation functional of Perdew, Burke, and Ernzerhof~\cite{Perdew1996}. The model consists of an H$_2$ molecule in a frozen p(2x2) Ag(111) surface unit cell. Further computational details can be found in the Supplemental Material (SM)~\cite{supplemental}. 

Fig.~\ref{fig-neb-friction} shows the EHP-induced relaxation rates along a minimum energy path of DC, resulting in H adsorption at adjacent fcc and hcp sites. The main features of tensorial electronic friction along these paths are: (1) Electronic friction is strongly mode dependent. As a result, friction along the bond stretch $d$ and the azimuthal and polar angles ($\theta$ and $\phi$) is considerably larger than along the center of mass translations of the molecule ($X$, $Y$, $Z$). (2) The relaxation rate along the bond stretch $d$ peaks at the transition state and is three times larger than the next largest component. This is in agreement with findings of Luntz \emph{et al.} for H$_2$ on Cu(111).~\cite{Luntz2009} Both of these findings stand in contrast to the description of electronic friction by LDFA, where the relaxation rate is an isotropic function of the embedding substrate density at the position of the adsorbate atoms. LDFA relaxation rates monotonically increase as the H atoms approach the substrate and they are identical in all directions (see SM~\cite{supplemental}). 

\begin{figure*}[htbp]
 \centering\includegraphics[width=2.00\columnwidth]{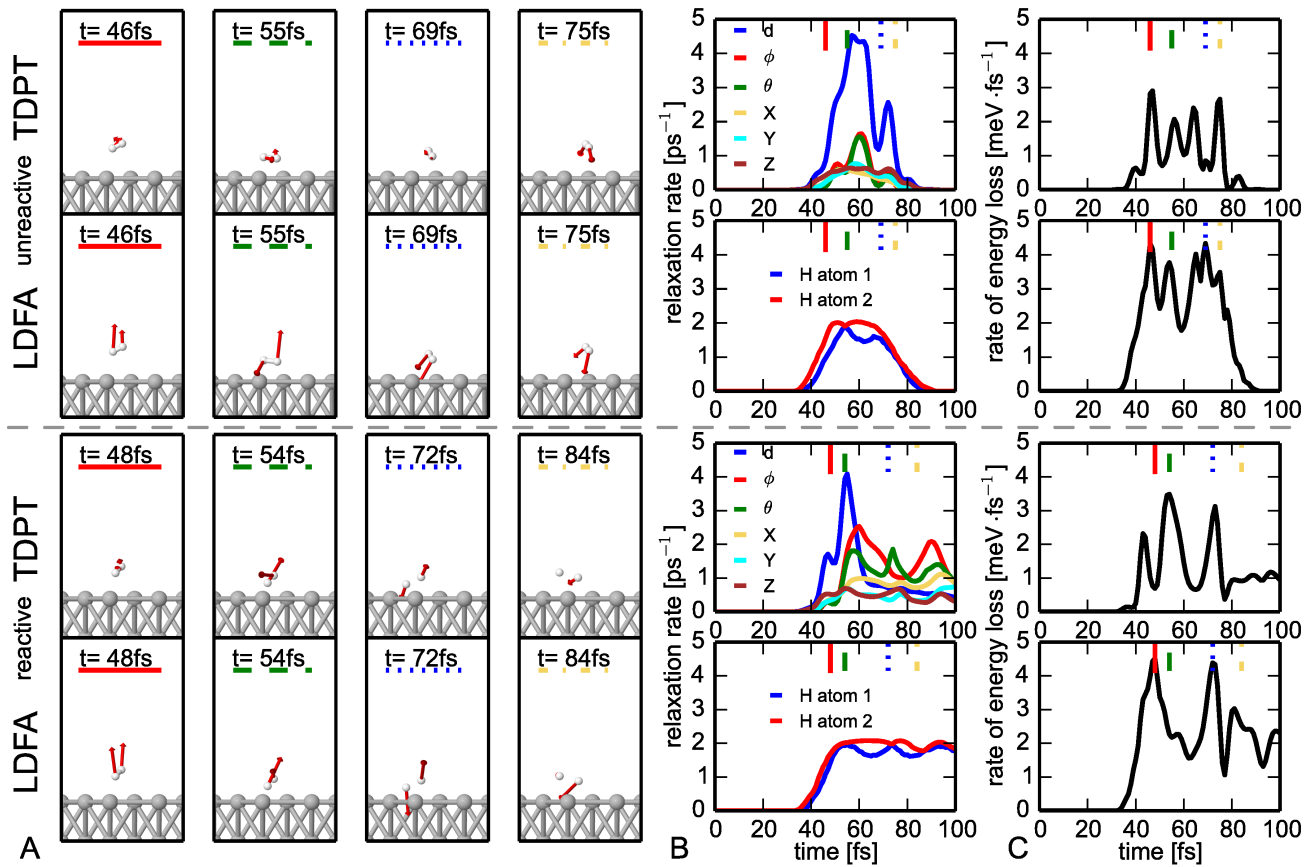}
 \caption{\label{fig-traj} (A) Snapshots at different times for an unreactive (top) and a reactive (bottom) trajectory. The incidence energy is 1.8 eV. Red arrows indicate the frictional force according to the second term in eq.~\ref{eq-langevin}. (B) Relaxation rate in ps$^{-1}$ projected along the internal coordinates defined in Fig.~\ref{fig-neb-friction}(C). (C) Rate of energy loss in meV/fs along the trajectory. Vertical lines indicate the snapshots in time shown in panel A.}
\end{figure*}

Stronger friction at the transition state is well known~\cite{Luntz2009} and readily understandable as electronic character changes dramatically during bond breaking. Nevertheless, the computational costs of TDPT-based friction models in full-dimensional MD have prevented further investigations of frictional effects during DC in the past. Our recent TDPT implementation allows us to overcome this computational constraint.~\cite{Maurer2016a} By explicitly simulating the reaction dynamics of H$_2$ on Ag(111) using MD, MD with electronic friction (MDEF) based on LDFA,~\cite{Juaristi2008,Li1992,Novko2015,Novko2016, Rittmeyer2015} and MDEF based on TDPT friction, we study the frictional energy loss of H$_2$ DC on Ag(111).~\cite{Jiang2014}
We have selected a total of 14 normal-incidence initial conditions varying in molecular orientation and impingement site with three different energies (see SM, Fig.~S1).~\cite{supplemental}
This includes 8 trajectories with a translational energy of 1.8~eV that exceeds the barrier for reaction (1.12~eV), 2 trajectories with a translational energy of 1.0~eV, which barely suffices to overcome the barrier, when accounting for vibrational zero point energy, and 4 vibrationally excited (v=1) trajectories with 0.6~eV translational energy. We integrate eq.~\ref{eq-langevin} with a precomputed PES~\cite{Jiang2014} and evaluate TDPT friction on-the-fly, including off-diagonal elements (see SM ~\cite{supplemental}). We assume a surface temperature of 0~K, effectively neglecting the third term in eq.~\ref{eq-langevin}. Inclusion of the random force term in eq. 1 to model non-zero temperatures would be  straightforward in our approach. However, in the current case, it would make it more difficult to draw conclusions from a small set of 14 trajectories.

Fig.~\ref{fig-traj} visualizes the MDEF results as given by two different friction models, the isotropic LDFA~\cite{Li1992,Juaristi2008} and the tensorial TDPT model.~\cite{Maurer2016a,Askerka2016} The figures represent two trajectories with 1.8~eV translational energy, one of which scatters from the surface, the other one dissociates. 
The frictional forces $F_{\mathrm{damp}}$ visualized in panel A correspond to the second term on the r.h.s. of eq.~\ref{eq-langevin}. The  nonadiabatic relaxation rates due to $\mathbf{\Lambda}(t)$ are given in Panel B of both figures. The corresponding energy loss along a given trajectory is presented in panel C and depends on, both, the velocity profile and the friction tensor at time $t$. 

In the case of non-dissociative molecular scattering, electronic friction is only non-zero during a short period of time (ca. 50~fs) in which the molecule-surface distance is small. In the LDFA method, friction is mostly dictated by the atom-surface distance and is isotropic in all directions. 
In the TDPT method, friction acts most strongly along the bond stretch coordinate with significantly more variation along the trajectory. However, the actual energy loss shows minor differences betwen the two methods and mostly differs in magnitude. We find an average integrated energy loss of 47~meV and 114~meV for the four scattering trajectories with 1.8~eV when calculated with TDPT and LDFA, respectively. This corresponds to an energy loss of only 2\% and 6\% of the total incidence energy. The reaction outcomes of these high-energy trajectories were not affected by electronic friction.

For successful chemisorption (bottom part of Fig.~\ref{fig-traj}), friction and energy loss show larger discrepancies between the two models. Upon adsorption, LDFA friction remains constant throughout the dissociation event, whereas TDPT-based friction exhibits more striking variation. The resulting energy-loss profiles are characterized by three spikes in energy loss along the dissociation path, which are weighted differently by the two friction methods. The first one (t=48~fs), corresponds to the initial adsorption event, the second one (t=54~fs) to the dissociation event, and the third (t=72~fs) corresponds to the onset of lateral diffusion. LDFA yields more energy loss for the first and third event, TDPT yields more energy loss for the dissociation event. 
Throughout the studied trajectories, we found that LDFA yields more energy loss perpendicular to the surface upon adsorption than TDPT. This can be seen from the corresponding force vectors in Fig.~\ref{fig-traj}. Overall, we can distinguish between two regions of the energy loss profiles: a highly structured region (t$<$80~fs) leading up to and including dissociation and a region (t$>$80~fs) with less variation in energy loss, corresponding to lateral atom diffusion. 
This suggests that isotropic friction models are appropriate for the description of single atom motion on metal surfaces,~\cite{Blanco-Rey2014} while they may have problems in representing molecular motion.~\cite{Luntz2009}

Individual reaction events with high incidence energy reveal significant structure in the energy loss profile of a chemisorption event. Nevertheless, we have not found significant changes in reaction probabilities or final energy distributions upon incorporation of friction. However, in order to unambiguously assess frictional effects on reaction probabilities and final energy distributions, extensive statistical sampling with thousands of independent trajectories would be necessary. This is, however, at the moment computationally challenging.

\begin{table}
\caption{\label{tab-outcome} Outcomes of individual trajectories calculatd with classical MD, MDEF based on LDFA, and MDEF based on TDPT. S refers to molecular scattering.}
\begin{center}
\begin{tabular}{ccccc} \hline\hline  \noalign{\vskip 1pt} 
\multicolumn{2}{c}{trajectory \#} & MD & MDEF(LDFA) & MDEF(TDPT) \\ \hline \noalign{\vskip 1pt} 
E=1.0~eV & 9 & DC & DC & DC \\ 
v=0 &10 & DC & S & S \\ \hline 
&  11 & S & DC & S \\
E=0.6~eV &  12 & S & DC & DC \\
v=1 &  13  & S & DC & DC \\
&  14 & DC & S & S \\ \hline\hline
\end{tabular}
\end{center}
\end{table}

Only when considering initial conditions closer to the dissociation barrier, we find changes in reaction outcomes and energy distributions (see Table~\ref{tab-outcome}). In many of the studied cases, the inclusion of electronic friction changes the reaction outcome from successful dissociation to scattering or \emph{vice versa}. The overall energy loss, in all cases is still around 5\% of the incidence energy and the total energy always remains sufficient to overcome the barrier in principal. Nevertheless, energy loss and directional steering along specific modes appears to be sufficient to change individual reaction outcomes. Interestingly, we find that absolute energy loss described by the two friction models is closer in magnitude for these trajectories, especially for vibrationally excited cases [see SM, Table~S~IV].~\cite{supplemental} This originates from the dominance of friction along internal modes in TDPT, yielding higher energy loss along rotations and vibrations. In one vibrationally excited case, we find final reaction outcomes that differed between the two friction models [see SM, Figs.~S19 and S20].~\cite{supplemental} Here, the hydrogen atoms dissociate onto adjacent hollow sites and immediately recombine. The translational energy loss after dissociation is significantly larger when employing LDFA friction than TDPT. As a result, hydrogen atoms cannot recombine and the dissociation remains successful. MD and MDEF(TDPT), on the other hand, both describe recombinative desorption, however with very different final rovibrational and translational energy distributions.

\begin{table}
\caption{\label{tab-energy} Initial (t=0) and final internal and translational energy contributions and EHP-induced energy loss for scattered trajectories in \%. Columns without results correspond to DC events. }
\begin{center}
\begin{tabular}{c|cc|cc|ccc|ccc} \hline\hline 
\# & \multicolumn{2}{c}{t=0}&  \multicolumn{2}{c}{MD} & \multicolumn{3}{c}{MDEF(LDFA)} & \multicolumn{3}{c}{MDEF(TDPT)} \\
& $E_{\mathrm{trans}}$ & $E_{\mathrm{int}}$ & $E_{\mathrm{trans}}$ & $E_{\mathrm{int}}$ & $E_{\mathrm{loss}}$ & $E_{\mathrm{trans}}$ & $E_{\mathrm{int}}$ &$E_{\mathrm{loss}}$ & $E_{\mathrm{trans}}$ & $E_{\mathrm{int}}$ \\ \hline 
1 & 87 & 13 & 69 & 31 & 7 & 64 & 30 & 3 & 67 & 30 \\
2 & 87 & 13 & 48 & 52 & 5 & 48 & 47 & 3 & 49 & 48 \\
4 & 87 & 13 & 76 & 24 & 7 & 71 & 22 & 3 & 75 & 22 \\
5 & 87 & 13 & 29 & 71 & 4 & 28 & 68 & 3 & 29 & 68 \\
 10 & 79 & 21 & - & - & 5 & 62 & 32 & 4 & 72 & 24 \\
11 & 44 & 56 & 63 & 37 & - & - & - & 6 & 92 & 2 \\ 
14 & 44 & 56 & - & - & 5 & 66 & 29 & 4 & 73 & 23 \\
\hline\hline
\end{tabular}
\end{center}
\end{table}

\begin{figure}
\centering\includegraphics{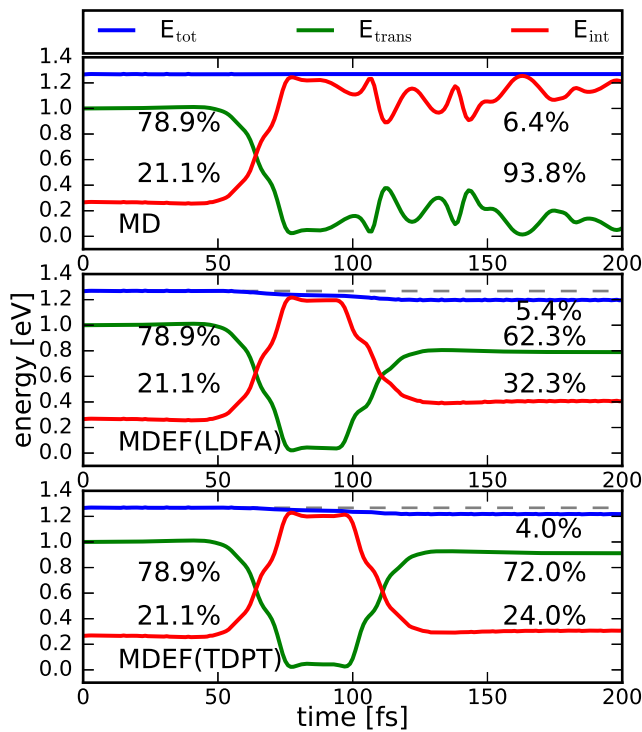}
\caption{\label{fig-edist} Time evolution of energy distribution for a trajectory with  E$_{\mathrm{trans}}$=0.6~eV and v=1. Shown are the translational energy, the internal molecular energy, and the total energy as calculated by MD, MDEF(LDFA), and MDEF(TDPT). }
\end{figure}

This brings us to the discussion of final energy distributions upon scattering. The energy distribution across different modes of a scattered molecule is an important experimental observable and can potentially expose the effects of electronic friction. Table~\ref{tab-energy} summarizes relative energy distributions of scattered trajectories before and after interaction with the surface. Coupling between vibration and rotations does not allow independent analysis and we collect them as internal energy ($E_{\mathrm{int}}$). The first four rows correspond to translationally hot molecules with 1.8 eV, where energy distributions with or without electronic friction are very similar. Energy loss occurs dominantly from translational energy, with LDFA friction yielding higher energy loss than TDPT. Qualitative differences between the two methods become much more obvious in the case of low incidence energy (\#10 in Table II) and vibrational excitation (\#11, \#14) where both friction models yield comparable energy loss. Fig.~\ref{fig-edist} visualizes the energy loss of \#10 as a function of time. Despite EHP-induced energy loss of 5\% and 4\% of the total energy, the two friction models differ significantly in the final amount of energy distributed over translation and rovibrational degrees of freedom, with more internal energy loss in the case of TDPT. The result is a difference in relative energy distributions between the two models of almost 10\%. In the case of the previously mentioned trajectory \# 11, where MD and MDEF(TDPT) both describe recombinative desorption, MD yields a final 2:1 energy distribution of translation and rovibration, MDEF yields desorption of an internally cold molecule. 

Nonadiabatic surface-adsorbate energy transfer is an essential aspect of gas-surface dynamics on metals and can be decisive in steering chemical reactions. Using an electronic friction model based on TDPT that accounts for the molecular structure of the adsorbate, we studied frictional effects in DC of H$_2$ on Ag(111). The overall EHP-induced energy loss of scattered trajectories only amounts to about 5\% of the total energy in both employed friction models. Nevertheless, for low translational energies and vibrationally excited molecules, significant frictional effects on directional steering and final energy distributions can be found. Measured dissociation probabilities might not necessarily reflect EHP-induced nonadiabatic effects, however, we expect effects on final energy distributions to be measurable in molecular beam scattering. Experimental evidence for this has recently been presented.~\cite{Shuai2017}
Obvious next steps include a statistically representative determination of the latter, as well as the study of temperature effects on scattering and reaction and the ability to control outcomes through varying electronic temperature.

\begin{acknowledgments}
RJM and JCT acknowledge financial support by the US Department of Energy - Basic Energy Science grant DE-FG02-05ER15677. H. G. thanks US National Science Foundation (CHE-1462019). B. J. thanks the National Natural Science Foundation of China (91645202 and 21573203). Computational support was provided by the HPC facilities operated by, and the staff of, the Yale Center for Research Computing.
\end{acknowledgments}


%

\end{document}